\documentclass{webofc}
\usepackage[varg]{txfonts}   

\def\ms{ms$^{-1}$}

\def\mj{M$_{\rm{J}}$}

\woctitle{Hot Planets and Cool Stars}
\begin{document}
\title{Status of the Calan-Hertfordshire Extrasolar Planet Search}
%
%

\author{James S. Jenkins\inst{1,2}\fnsep\thanks{\email{jjenkins@das.uchile.cl}} \and
        Hugh R. A. Jones\inst{2} \and
        Patricio Rojo\inst{1} \and
        Mikko Tuomi\inst{2,3} \and
        Matias I. Jones\inst{1,4} \and
        Felipe Murgas\inst{5,6} \and
        John R. Barnes\inst{2} \and
        Yakiv Pavlenko\inst{2,7} \and
        Oleksiy Ivanyuk\inst{7} \and
        Andr{\'e}s Jord{\'a}n\inst{8} \and
        Avril C. Day-Jones\inst{1,2} \and
        Maria-Teresa Ruiz\inst{1} \and
        David J. Pinfield\inst{2}
}

\institute{Departamento de Astronomia, Universidad de Chile, Camino el Observatorio 1515, Las Condes, Santiago, Chile, Casilla 36-D
\and
Center for Astrophysics, University of Hertfordshire, College Lane Campus, Hatfield, Hertfordshire, UK, AL10 9AB
\and
University of Turku, Tuorla Observatory, Department of Physics and Astronomy, V\"ais\"al\"antie 20, FI-21500, Piikki\"o, Finland
\and
European Southern Observatory, Casilla 19001, Santiago, Chile
\and
Instituto de Astrof\'isica de Canarias, Via Lactea, E38205, La Laguna, Tenerife, Spain
\and
Departamento de Astrof\'{i}sica, Universidad de La Laguna (ULL),  E-38206 La Laguna, Tenerife, Spain
\and
Main Astronomical Observatory of National Academy of Sciences of Ukraine, 27 Zabolotnoho, Kyiv 127, 03680, Ukraine
\and
Departamento de Astronom\'ia y Astrof\'isica, Pontificia Universidad Cat\'olica de Chile, 7820436 Macul, Santiago, Chile
          }

\abstract{In these proceedings we give a status update of the Calan-Hertfordshire Extrasolar 
Planet Search, an international collaboration led from Chile that aims to discover more planets around 
super metal-rich and Sun-like stars, and then follow these up with precision photometry to hunt for 
new bright transit planets.  We highlight some results from this program, including exoplanet and brown 
dwarf discoveries, and a possible correlation between metallicity and planetary minimum mass at the lowest 
planetary masses detectable.  Finally we discuss the short-term and long-term future pathways this program 
can take.
}
\maketitle
\section{Introduction}\label{intro}

Precision radial velocity studies of the nearest stars have given rise to a new branch of astrophysics.  The study of 
planets orbiting stars other than the Sun (aka exoplanets) has revolutionised the way we view the Universe and our 
understanding of planets and planetary systems, including our own Solar System.  

The census of planetary systems within around 50~pc or so from the Sun that host gas giants has been well studied out to orbits approaching that 
of Jupiter (\cite{jones10}; \cite{boisse11}).  These discoveries have led to our knowledge of planet formation and evolution maturing 
at an ever accelerated rate and have shown that core accretion and planet migration appear to be the dominant mechanisms that 
sculpt these systems (\cite{lin86}).  

The earliest observational correlation between the presence of planets and a measurable quantity was found by \cite{gonzalez97} 
who noticed that the three gas giant planets known at that time were found to orbit stars rich in metals in comparison to our Sun.  
This correlation has been extensively studied and recent works have found that the probability of planet formation is a rising power law 
with the metallicity of the host star (\cite{fischer06}).  Metallicity in this sense is characterised by the iron abundance ([Fe/H]) however 
other metallic elements possibly correlate also with the presence of planets, elements such as lithium (\cite{israelian09}), oxygen, chromium, 
and yttrium (\cite{bond08}), etc.

Given that stellar metallicity plays a key role in the formation of planets and therefore their final configurations, studying sub-samples of 
stars like the Sun as a function of metallicity can allow one to target specific types of planets.  With this in mind we started a planet search 
project on the HARPS spectrograph, as part of a collaboration between the Universidad de Chile and the University of Hertfordshire, that 
aims to discover more gas giants around super metal-rich stars in the southern hemisphere and follow them up with Chilean facilities to 
test if they transit their parent stars.  In this conference proceedings we discuss some of the recent findings from our Calan-Hertfordshire 
Extrasolar Planet Search project and future paths the project will take.

\section{The Calan-Hertfordshire Extrasolar Planet Search}\label{cheps}

\subsection{Sample Selection}\label{sample}

The Calan-Hertfordshire Extrasolar Planet Search (aka CHEPS) is a metal-biased planet search project, focusing on main sequence and 
subgiant Sun-like stars with super solar metallicities.  The initial sample of stars were selected from the Hipparcos catalogue (\cite{perryman97}) 
to be southern ($\delta~\le$~90$^{\rm{o}}$), to have $B-V$ colours in the range 0.5-0.9, bracketing the late F to early K star regime, and to 
have Johnson $V$ magnitudes between 7.5-9.5.  This magnitude range was selected since all such stars brighter than 7.5 in $V$ were already 
being observed by other programs, and the upper magnitude limit of 9.5 was set to ensure the stars were bright enough that if any transit was 
detected the planet could easily be followed-up for secondary eclipse measurements and transmission spectroscopy.

Once this selection was made we ensured that no star was classed as having a stellar companion within 2$''$ from Hipparcos, along with 
selecting non-varying stars, and spectral classes of VI or V.  From this selection we originally followed up the first 350 using the FEROS 
instrument at la Silla observatory in Chile and measured precision metallicities, along with precise chromospheric activities.  The methodology 
for measuring the activities followed that in \cite{jenkins06}, however the method for measuring the metallicities was a new method that 
focused on a small number ($\sim$30) of Fe~\sc i\rm ~lines that appeared unblended by any neighbouring atomic lines, allowing us to 
determine the spectral synthesis model that best describes the data.  This technique produced metallicities with precisions between 0.03-0.10~dex 
(\cite{jenkins08}) and gave rise to 105 inactive (logR$'_{\rm{HK}}~\le~$-4.5~dex) and metal-rich ([Fe/H]$~\ge~$+0.1~dex) target stars to observe 
with HARPS in the hunt for new planetary systems. 

Since this work we have observed a further 600 of the Hipparcos selected stars with FEROS and have published their activities, kinematics, and 
rotational velocities (\cite{jenkins11}).  The metallicities and chemical abundances for these stars are still being processed and using our updated 
method for measuring precision abundances, microturbulent velocities, and rotational velocities (\cite{pavlenko12}).

\subsection{Early Discoveries}\label{bds}

\subsubsection{Brown Dwarf in the Desert}

\begin{figure}
\centering
\includegraphics[width=10cm,clip]{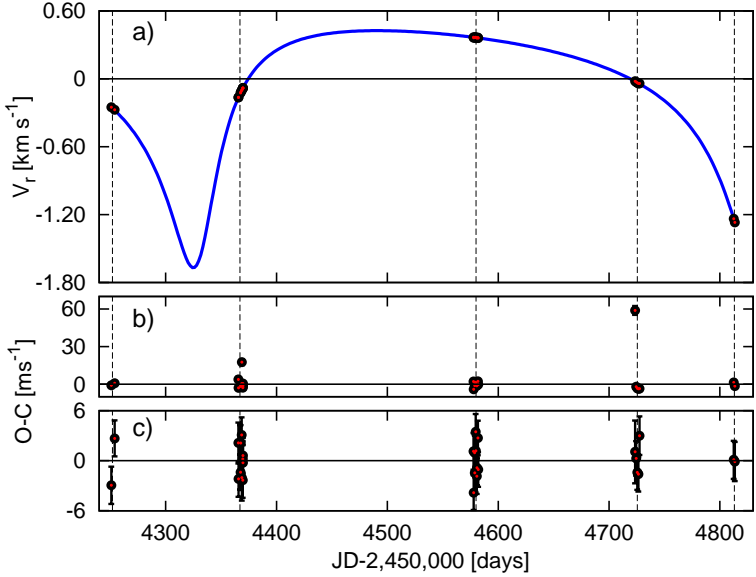}
\caption{Panel (a) shows the best fit Keplerian solution to the Doppler points for HD191760 with BVS correction.  
Panel (b) shows the residuals without BVS correction and panel (c) shows the residuals after correction.}
\label{hd191760_rv}   
\end{figure}

The first radial velocity results from the CHEPS program were published in 2009 (\cite{jenkins09}) and included the discovery of a 
brown dwarf in the desert orbiting the Sun-like star HD191760.  In Fig.~\ref{hd191760_rv} we show the radial velocity solution for 
this system, along with the residuals to the fit.  

In panel a (top) of the figure the Doppler velocities are in red and the best fit Keplerian solution in blue.  The data span five separate observing 
epochs, consisting of a total number of individual measurements of 29.  The Keplerian fit describes the data well, but only after a bisector 
correction was applied to the data.  This Bisector Velocity Span (BVS) correction was applied as in \cite{migaszewski10} and was found to 
correlate linearly with the radial velocities residuals to the best fit before applying any correction, and was found to have an index $\alpha$ of 0.697$\pm$0.064.  

Panels b and c (middle and lower) we see firsthand the effects of the BVS correction to the velocities.  In b we see the residuals to the best fit before 
the BVS correction has been applied.  Two clear outliers in the residuals are present, which were data points that were observed in bad weather 
conditions and hence had low S/N.  Once the BVS correction has been applied to the velocities we find no outliers in the residuals after refitting 
the best Keplerian model, as is seen in c.  The final solution here indicates a brown dwarf with a minimum mass of 38.17\mj ~orbits HD191760 
with an orbital period of 505~days and eccentricity of 0.63.

\subsubsection{Gas Giant Exoplanets}

\begin{figure}
\centering
\includegraphics[width=6cm,clip]{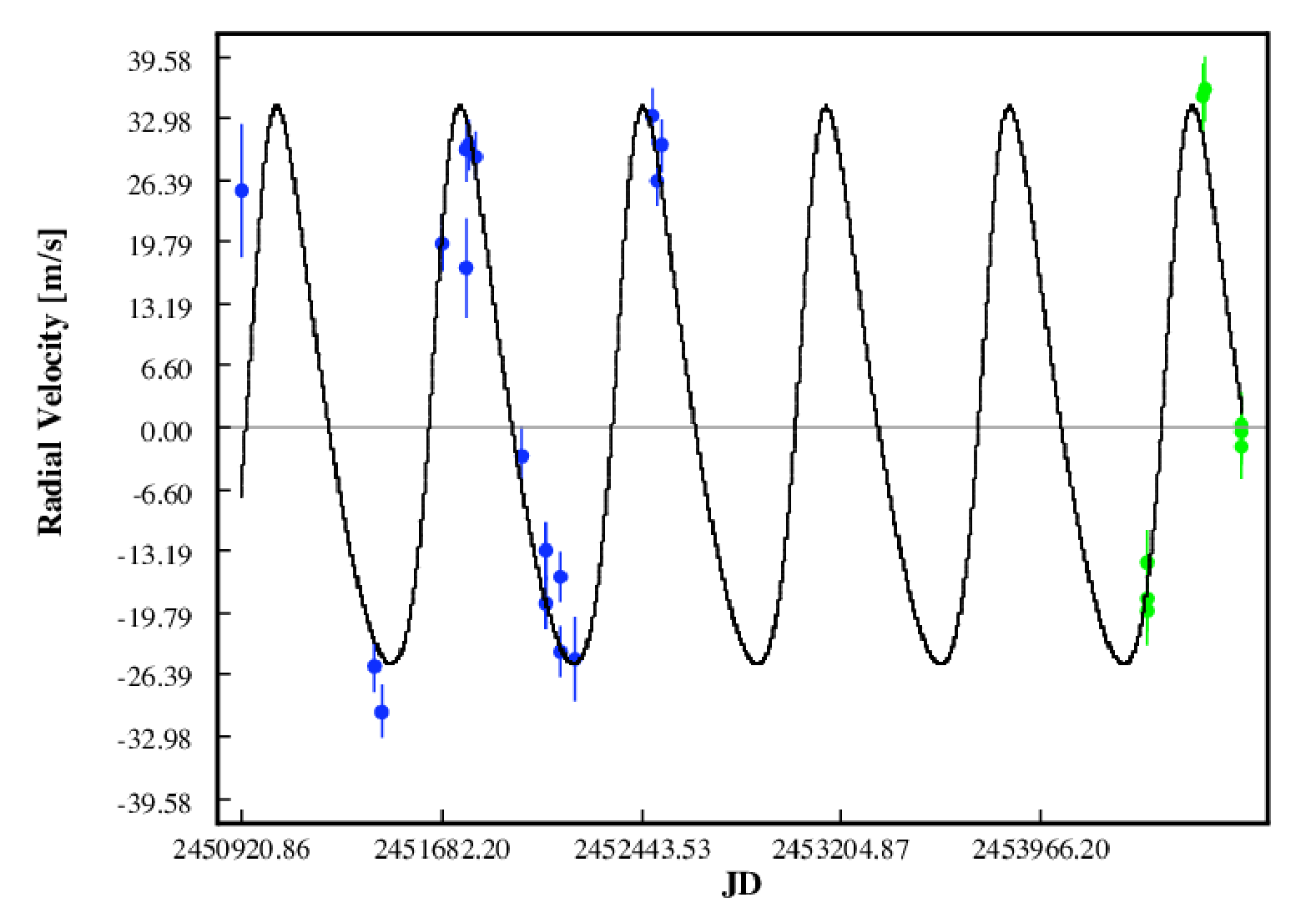}
\includegraphics[width=6cm,clip]{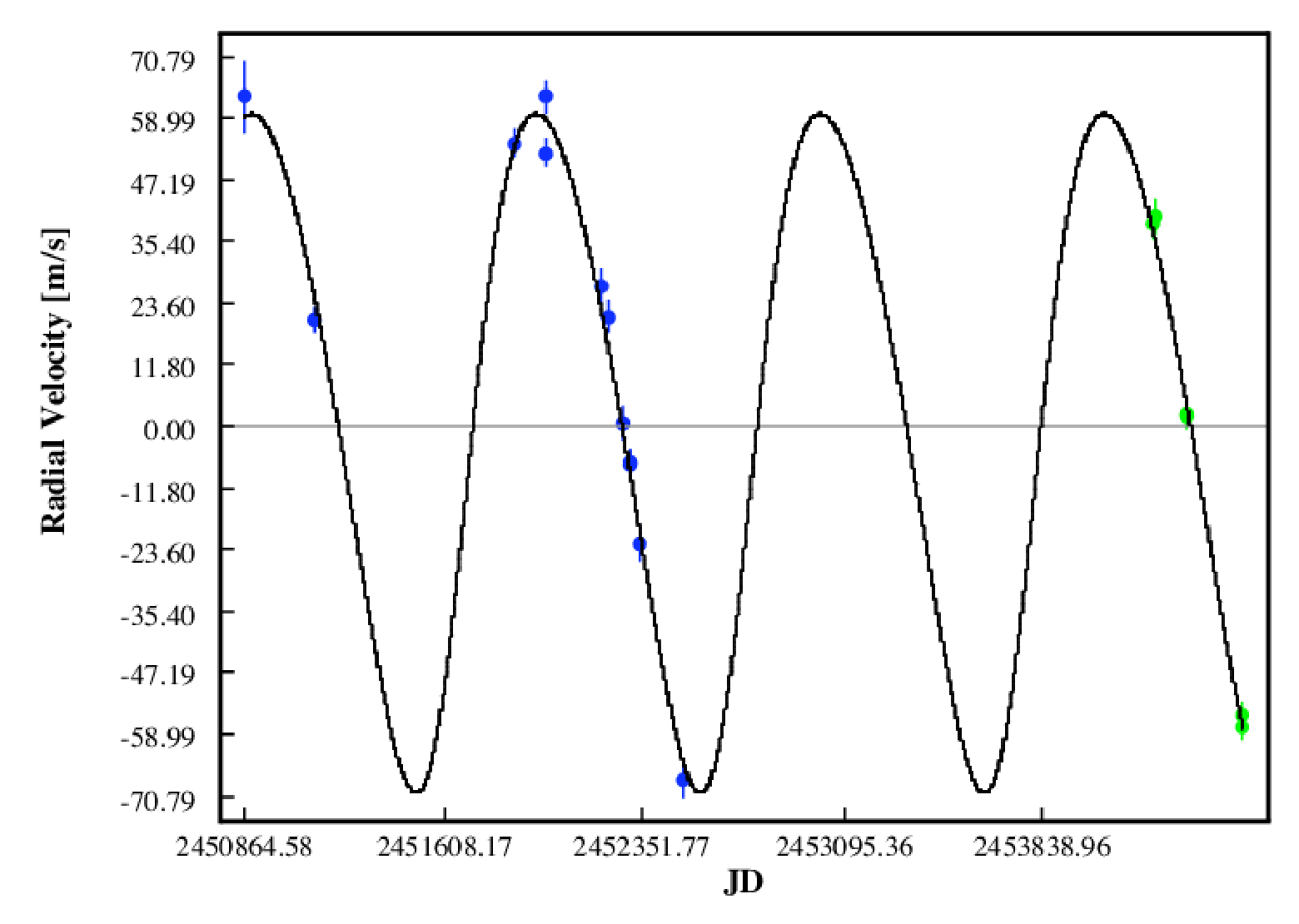}
\includegraphics[width=6cm,clip]{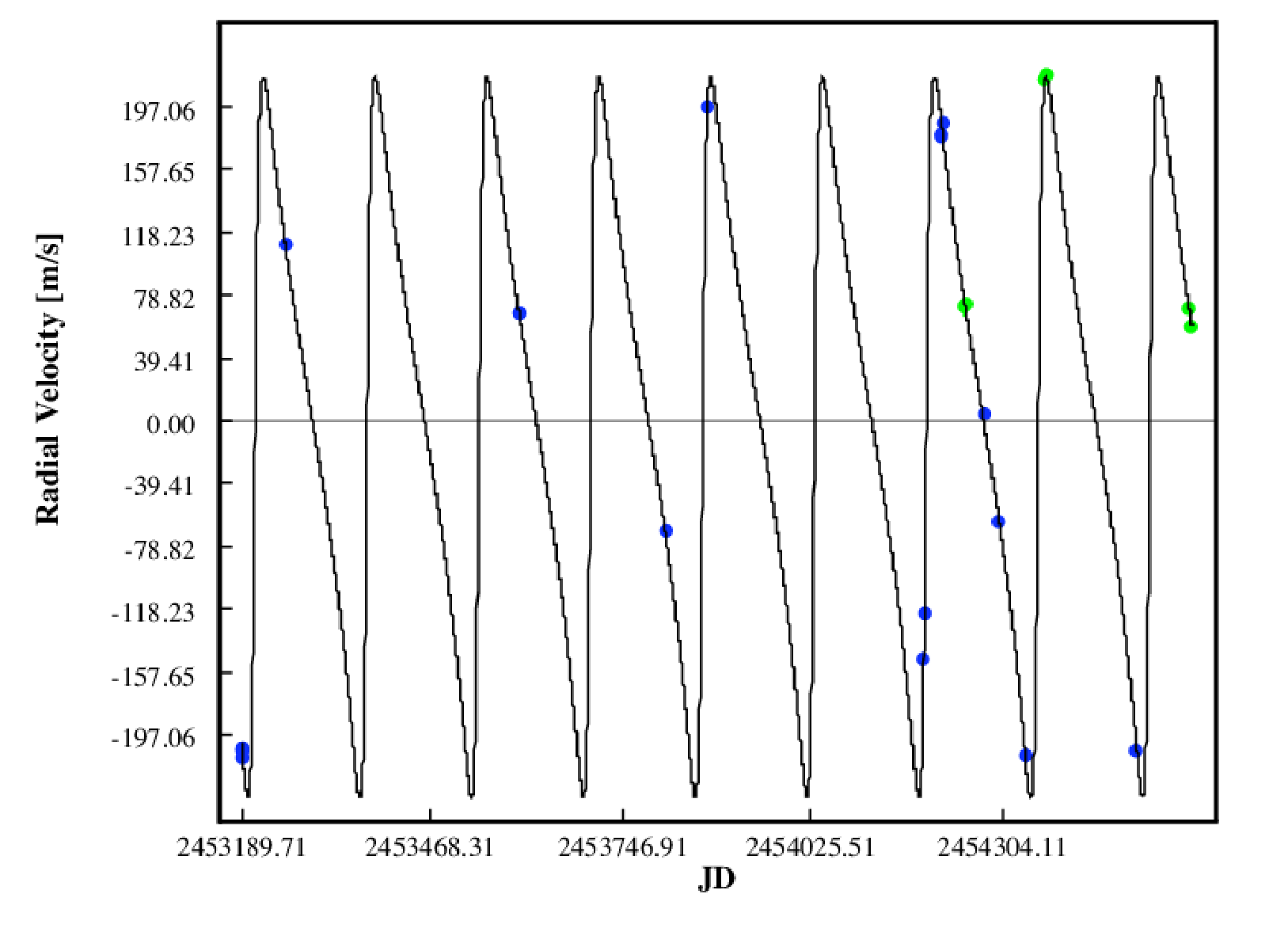}
\caption{The radial-velocity Keplerian fits to the stars HD48265 (top left), HD143361 (top right) and HD154672 (bottom).  The HARPS radial 
velocities are plotted in green and the Magellan radial velocities are shown in blue.}
\label{rvs}   
\end{figure}

Along with the discovery of the brown dwarf companion we also found gas giant planets orbiting three of our other target stars (see \cite{jenkins09}).
The radial velocity curves for the stars HD48265 (top left), HD143361 (top right), and HD154672 (bottom) are shown in Fig.~\ref{rvs}.  All three gas giant 
planets have masses above a Jupiter-mass, all appear to have measurable eccentricity, and they have orbital periods ranging from 160 days out to 
1060~days.  Our velocities from HARPS complement the radial velocity data from two other works from Magellan (\cite{lopez-morales08}; \cite{minniti09}).

\subsection{Latest Radial Velocities}

Since the 2009 work we have acquired significantly more radial velocity data and also expanded the program from the HARPS instrument 
to also include the Coralie spectrograph.  Coralie was the precursor instrument to HARPS and is similar in design except the spectrograph 
is not vaccuum sealed and therefore not controlled in pressure (see \cite{udry00}).  However, Coralie can yield precision radial velocities for 
stars in the CHEPS sample at the level of around 7.5\ms ~(\cite{jenkins13}).

\begin{figure}
\centering
\includegraphics[width=6cm,angle=90,clip]{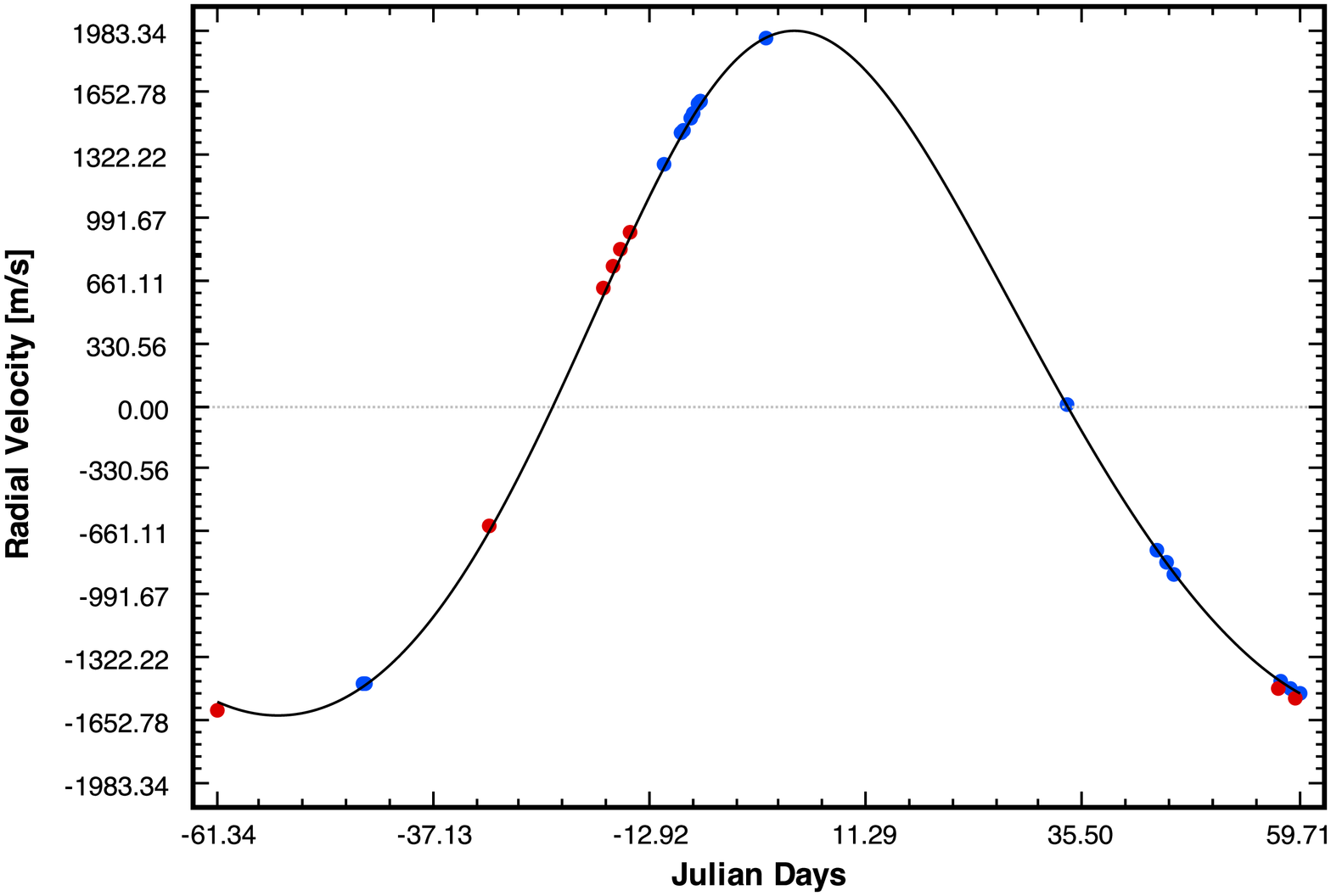}
\includegraphics[width=6cm,angle=90,clip]{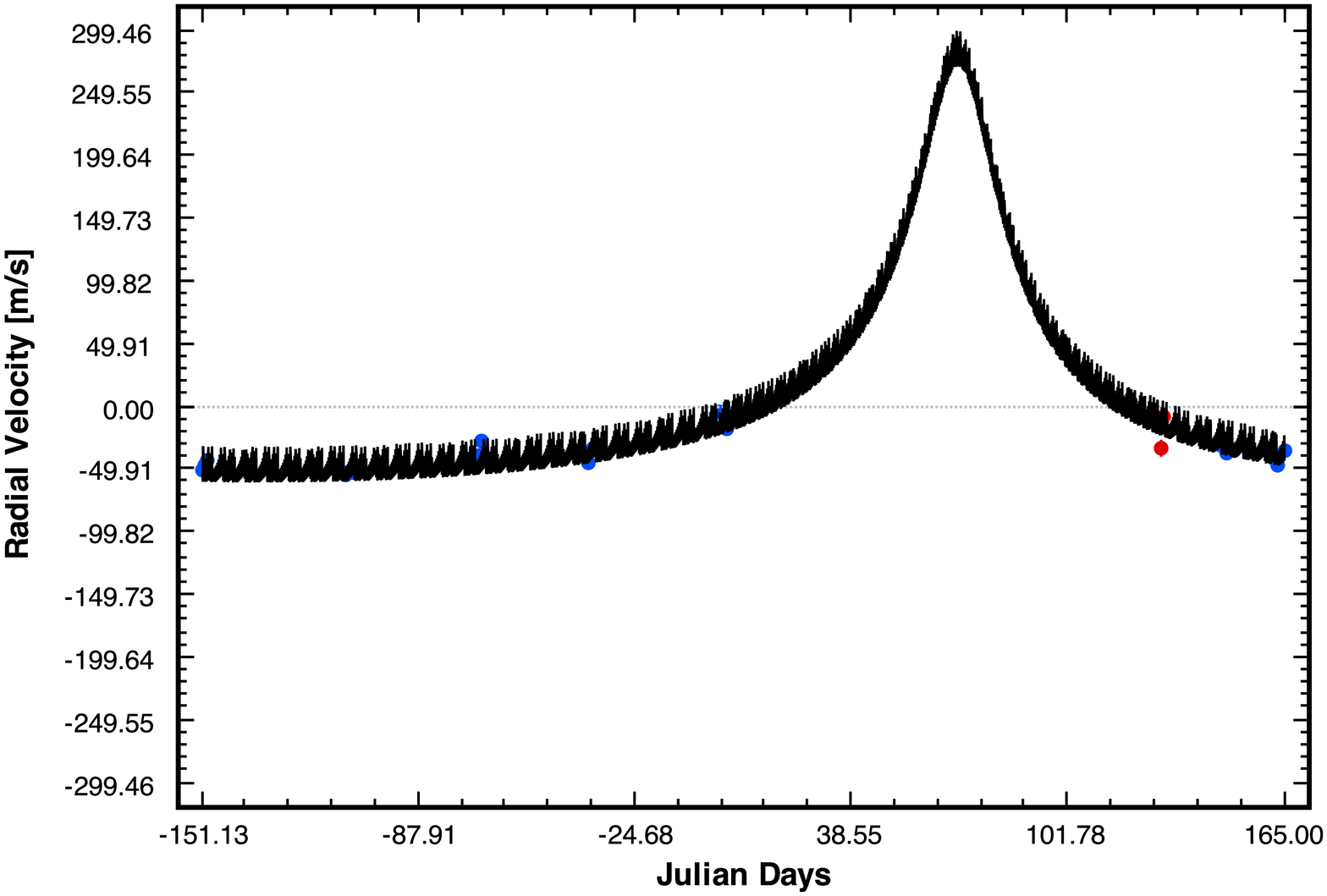}
\includegraphics[width=6cm,angle=90,clip]{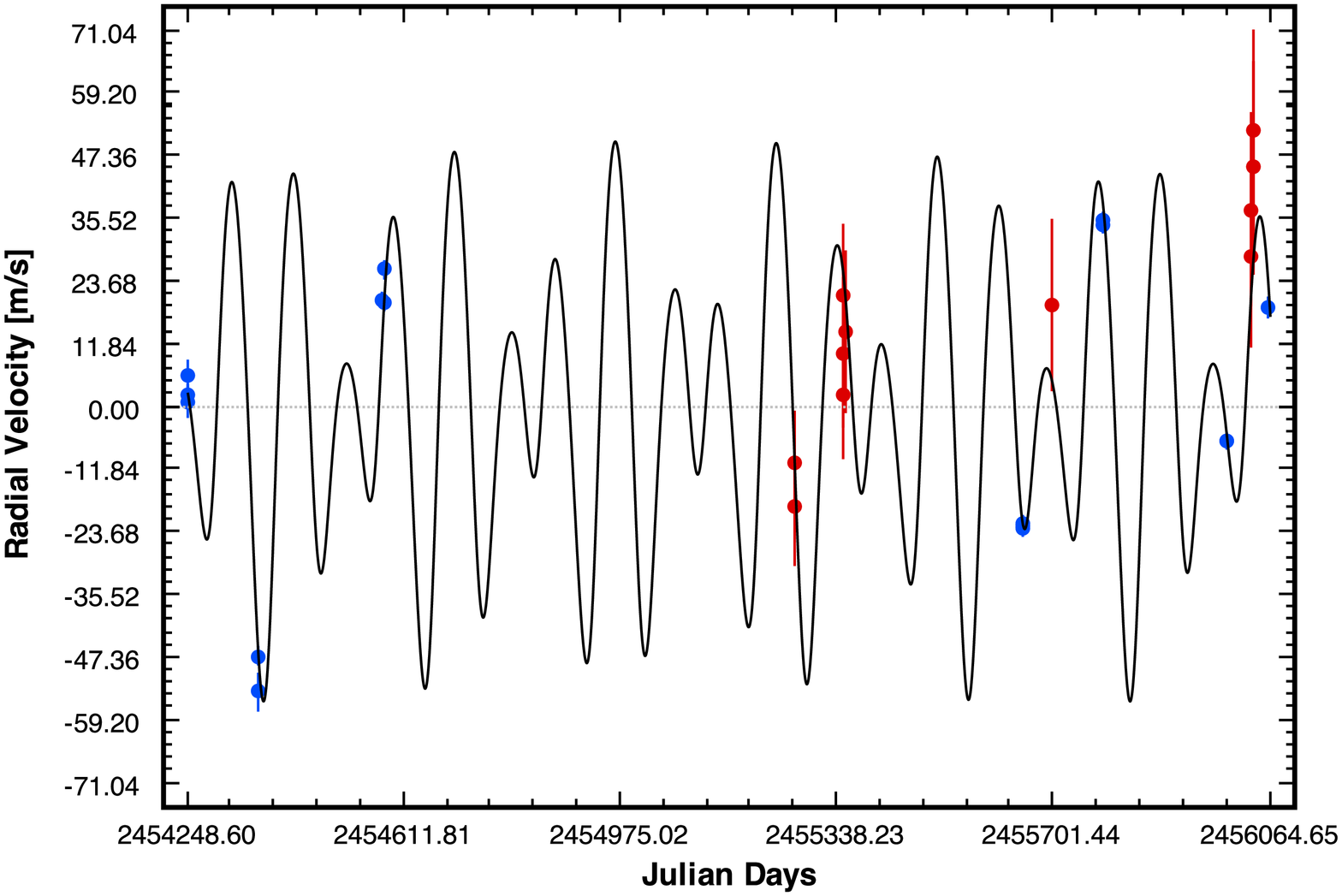}
\caption{A trio of emerging signals from the CHEPS.  The red data comes from Coralie and the blue data comes from HARPS.}
\label{signals}   
\end{figure}

In Fig.~\ref{signals} we show a small sample of the emerging signals from the CHEPS.  The three systems show a new brown dwarf companion in the 
brown dwarf desert (top), and two possible multi-planet systems (middle and bottom).  These three velocity curves give a nice representation of 
what is emerging from the data where we have acquired a significant amount of Doppler points from both HARPS and Coralie. 

\begin{figure}
\centering
\includegraphics[width=5cm,angle=-90,clip]{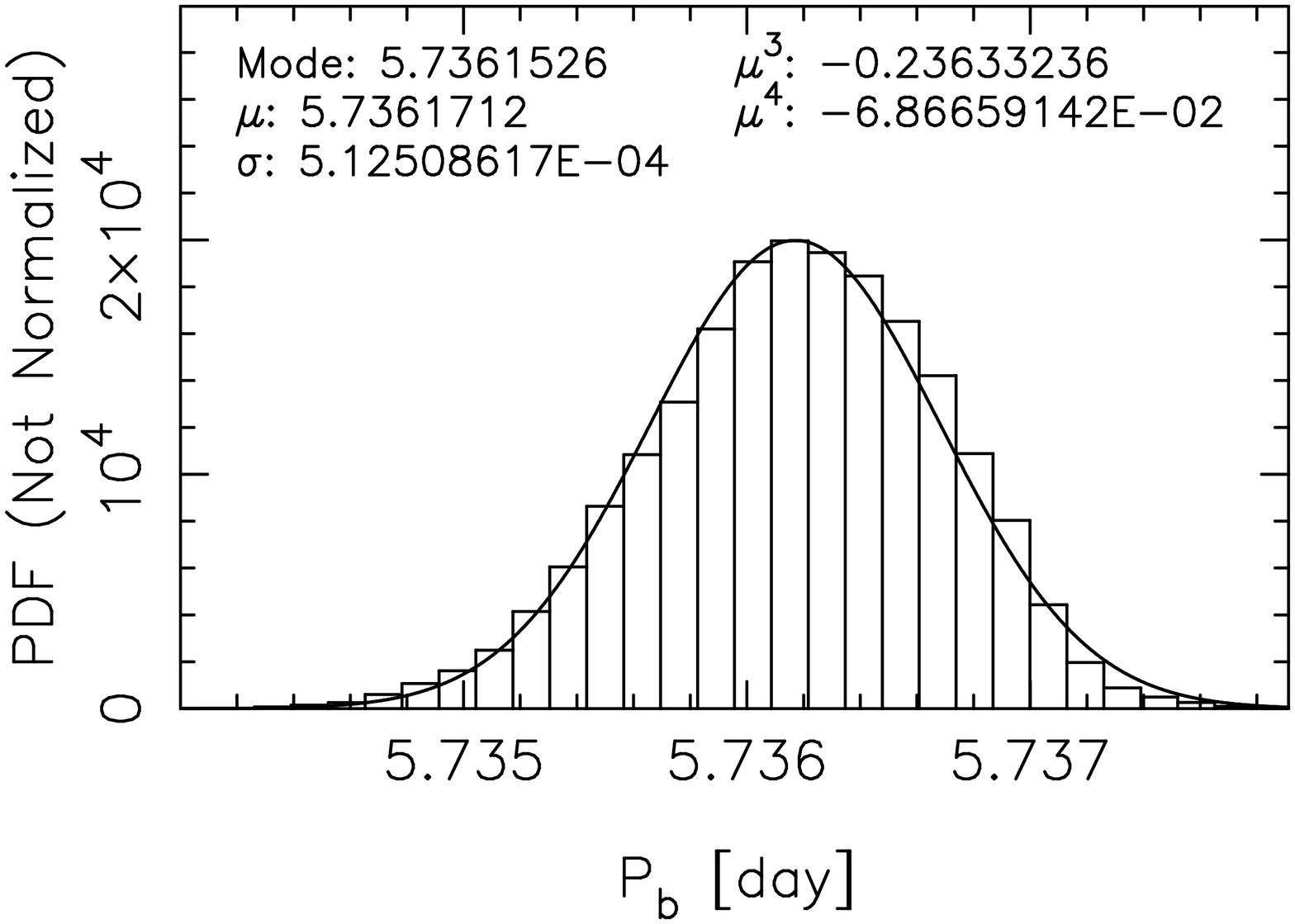}
\includegraphics[width=5cm,angle=-90,clip]{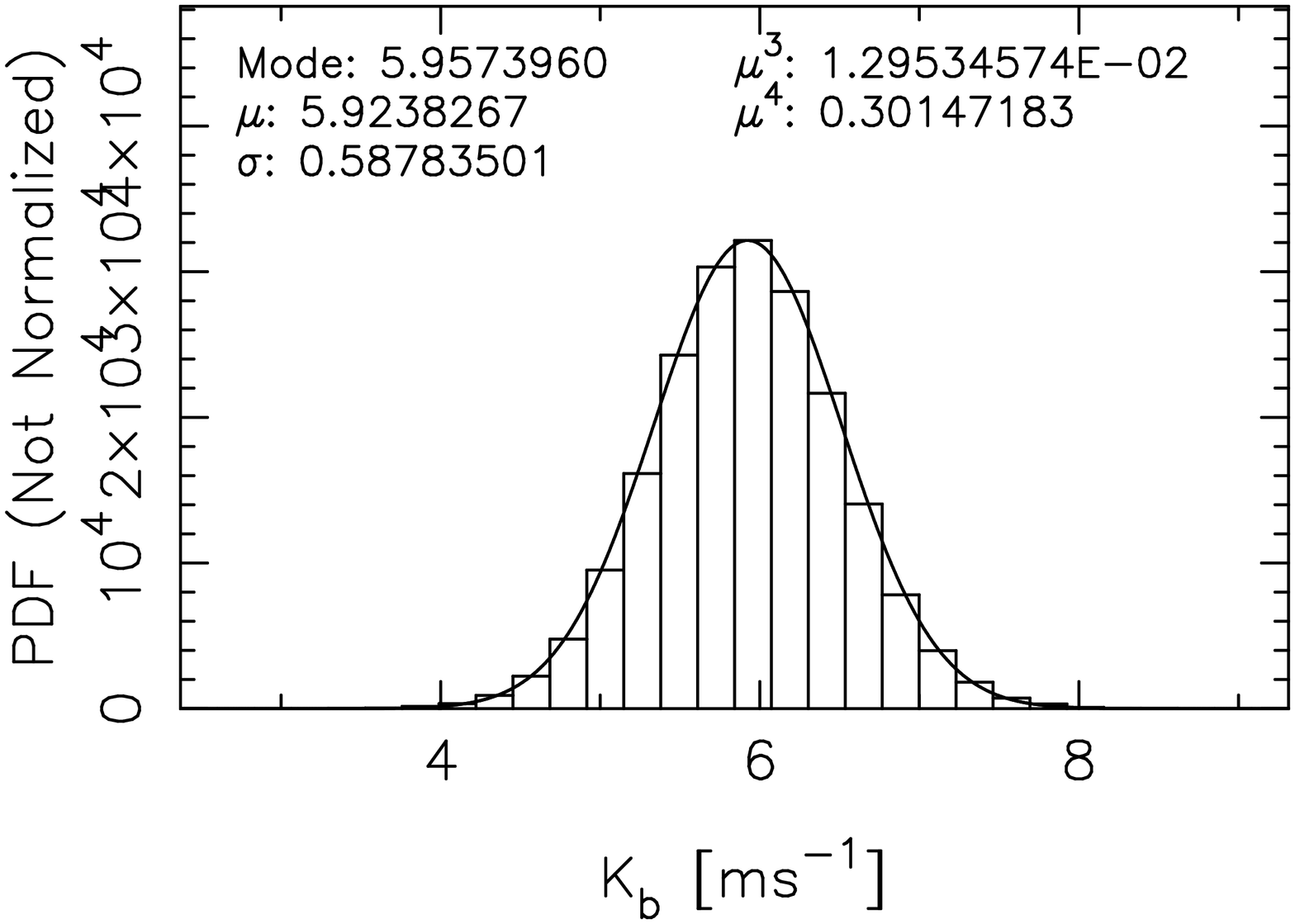}
\includegraphics[width=5cm,angle=-90,clip]{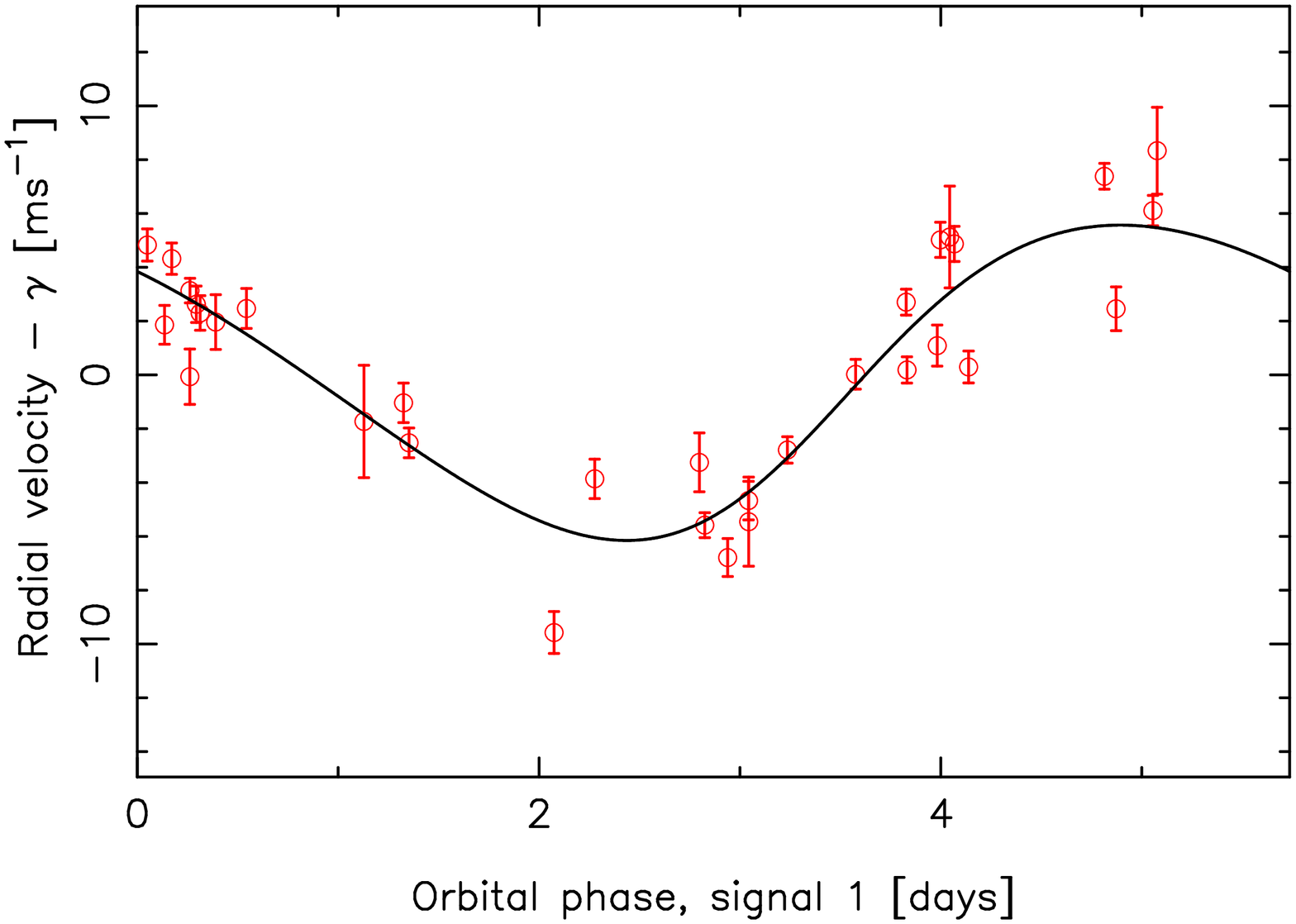}
\caption{Bayesian posterior probability densities for the radial velocity data set of HD77338 are shown for both signal period (top left) and semi-amplitude (top right).  
The lower panel shows the phase folded best fit Keplerian signal to the data.}
\label{planet}   
\end{figure}

In addition to these emerging candidates we have recently published a hot Uranus-mass planet candidate to the most metal-rich single star known 
to host a sub Neptune-mass planet, HD77338 (\cite{jenkins12}).  In Fig.~\ref{planet} we show the Bayesian posterior probability density distributions for the signal 
period and the semi-amplitude and also show the phase folded radial velocities.  

\begin{figure}
\centering
\includegraphics[width=6cm,angle=90,clip]{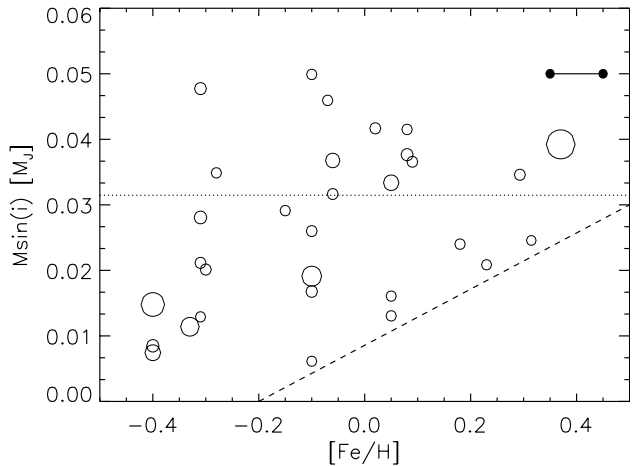}
\caption{Metallicity against minimum mass for all radial velocity detected
planets with a minimum mass less than that of Neptune.
The horizontal dotted line highlights the canonical boundary where runaway
gas accretion onto the growing core is expected to occur. The dashed line marks
the lower boundary in mass and the data points have been scaled in size by their
orbital period. The filled circles connected by a straight line show the position
of HD77338$b$.}
\label{metals}   
\end{figure}

Adding this planetary companion to the list of known sub Neptune-mass planets helps to uncover a possible correlation between metallicity and minimum 
planetary mass, whereby the lowest-mass planets yet detectable are under-abundant orbiting the most metal-rich stars.  Such a correlation can be placed in the 
framework of core accretion planet formation where proto-planetary cores acquire more mass in a given time interval than cores in disks with a lower 
abundance of metals.  Therefore, we find a planetary desert in the super metal-rich regime compared to the sub-solar metallicity regime (Fig.~\ref{metals}).

\section{Summary and Future Prospects}

In this conference proceedings we briefly give a status update from the Calan-Hertfordshire Extrasolar Planet Search.  We highlight some of the results 
from this project, both published and yet to be published.  These results include discovered brown dwarf companions to super metal-rich stars in 
the brown dwarf desert region of parameter space, gas giant planet discoveries, possible multi-planet systems, low mass planet candidates, and 
a correlation between metallicity and planetary minimum mass for sub-Neptune mass planets.

The study of extrasolar planets is following a number of paths at the present time, from attempts to discover the first Earth-like planets in the 
habitable zones of their parent stars (e.g. \cite{anglada-escude12}; \cite{barnes12}; \cite{tuomi12}), to more detailed studies of the atmospheres of 
these planets (\cite{swain10}; \cite{bean}), along with studying planets in different stellar environments (e.g. \cite{jones11}) to name a few.  However, in the near future we 
aim to expand the CHEPS sample using our full database of precise metallicities and stellar activities and hunt for more gas giant planets with a high 
probability to transit their host stars.  In the shorter term we shall publish the next clutch of planet and brown dwarf companions from the sample, 
and then complete a Bayesian study of the sample signal fractions, to further constrain the frequency of planet formation around metal-rich stars.

\end{document}